# The Sounds of Cyber Threats

**Akbar Siami Namin**
Computer Science Department
Texas Tech University
Lubbock, TX 79409, USA
akbar.namin@ttu.edu

**Keith S. Jones**
Psychology Department
Texas Tech University
Lubbock, TX 79409, USA
keith.s.jones@ttu.edu

**Rattikorn Hewett**
Computer Science Department
Texas Tech University
Lubbock, TX 79409, USA
rattikorn.hewett@ttu.edu

**Rona Pogrund**
College of Education
Texas Tech University
Texas School for the Blind and Visually Impaired (TSBVI)
Austin, TX, 78756, USA
rona.pogrund@ttu.edu

## Abstract
The Internet enables users to access vast resources, but it can also expose users to harmful cyber-attacks. This paper investigates human factors issues concerning the use of sounds in a cyber-security domain. It describes a methodology, referred to as sonification, to effectively design and develop auditory cyber-security threat indicators to warn users about cyber-attacks. A case study is presented, along with the results, of various types of usability testing with a number of Internet users who are visually impaired. The paper concludes with a discussion of future steps to enhance this work.

## ACM Classification Keywords
H.5.2. **[User Interfaces]**: User-centered design, user interface management systems

## Introduction
The Internet increasingly plays important roles in our day-to-day activities. While it improves the quality of life, it also makes users vulnerable to various cyber-threats such as phishing attacks that lure users to malicious Websites. Given the prevalence of and risks pertinent to cyber-attacks, it is imperative that users be informed when they are being attacked. Many technologies and visual cues have been developed to help detect and determine the risks associated with cyber threats. While these cues are useful, the main drawback of using *only* visual cues for cyber attack alert is that they not always available to all users, in particular, those with visual impairments. The use of sounds could be a good complement to visual cues.

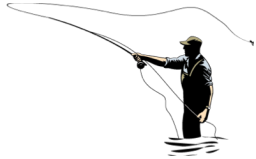

Figure 1. Casting a fishing rod - The sonification for phishing attack.

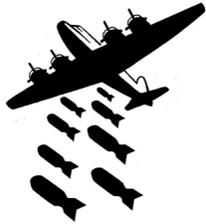

Figure 2. Dropping bombs - The sonification for malvertising.

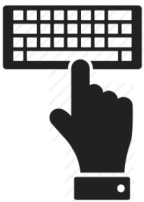

Figure 3. Typing on a keyboard - The sonification for form-filling.

The goals of this research were to investigate (1) how practical it is to convey security alerts to the targeted victims through sonification, and (2) whether Internet users find these sounds to be accessible and easy to use. This paper presents a methodology for sonifying cyber-security threats using non-speech sounds, earcons, in order not to further complicate the already sound-intensive screen reader users [6, 7, 8]. The focus was on sonifying three security threats and cues: *phishing*, *malvertising*, and *form-filling (or typing sensitive information)* into a form. Basic steps and results are briefly described below. A detailed and context-dependent report is published in [10].

## Sound-Featured Design: Issues & Approach

Designing effective computer-human interfaces to warn users, particularly those with visual impairments, in the context of cyber-attacks, faces several challenges. First, while sonifying warnings is well studied, almost all do not apply to cyber-attacks, which can take many forms, many of which are difficult to understand and articulate and are even unknown to novice users. The design must be able to convey the warning at different levels of danger to users with diverse backgrounds. Second, users with visual impairments use screen readers, which auditorily present information that normally would be graphically displayed on a computer screen (e.g., conveying text via speech). Microsoft Internet Explorer with a screen reader such as Job Access with Speech (JAWS) speaks the elements of the computer's interface aloud. The user hears synthesized speech as she navigates a document or the Internet on the computer screen. Human linear-like perception of sounds is limited compared to the nature of visual perceptions. Finally, the visual cues that individuals without visual impairments use to detect cyber-attacks can be lost when screen readers translate the system's interface into a verbal description. It is difficult for visually impaired users to know that those around them can see their information being typed into an insecure field.

This research proposes an earcon-based sonification design. Three approaches to designing non-speech sounds or earcons include: the *representational*, *abstract,* and *semi-abstract* approaches [1, 4]. The representational approach uses natural sounds. For example, the threat of a security breach could play the sound of an old creaky door. The abstract approach uses sounds that are synthesized from basic sound components. Combining pitches with simple rhythm and pitch design can be used to represent complex information. The semi-abstract approach is a mixture of the two.

The design principle focused on user-centric information that uses sonification as a means to convey i) the "*concept*" or "*meaning*" of associated cyber-security threats. ii) the "*consequences*" of the threats for users. and iii) the "*actions*" that users should take in response to the current threat. There are many types of cyber-security threats [9]. The CIA (Confidentiality, Integrity, and Availability) organizes attacks based on whether they threaten the **C**onfidentiality of information and users, tamper with the **I**ntegrity of information, or hinder the **A**vailability of services. The selection of sonification approaches largely depends on the underlying context. Selecting appropriate sonification approaches also rely on the ease to remember, the types of audiences, the familiarity of sonifications, and the relevance to the events sonified. Similarly, the selection of sounds mainly depends on the type and severity of events being sonified [4]. For instance, an event with catastrophic consequences should have loud and high frequency sounds.

| Threat | Sonification | Correctly Identified | Average Pleasantness | Average Urgency | Average Conspicuity | Rated Best | Correctly Remembered |
|---|---|---|---|---|---|---|---|
| Phishing | - Casting a fishing reel | 80% | 2.4 | 3.6 | 4.4 | 60% | 40% |
|  | - Breaking Glass | 0% | 2.8 | 4 | 4.2 | 40% | 60% |
|  | - Opening a Rusty Door | 0% | 2.4 | 2.8 | 3 | 0% | 60% |
| Malvertising | - Dropping a bomb | 40% | 3 | 4.2 | 5 | 20% | 80% |
|  | - Pouring Water | 20% | 3.2 | 2.8 | 4.4 | 0% | 40% |
|  | - Sounding a Siren | 60% | 2.6 | 4.6 | 4.8 | 80% | 60% |
| Form-Filling | - Typing on a keyboard | 40% | 3.4 | 3 | 2.8 | 100% | 80% |
|  | - Bubbling Water | 20% | 3.4 | 3 | 4 | 0% | 80% |
|  | - Playing a Slot Machine | 20% | 3.2 | 3 | 4.4 | 5 | 0% | 80% |

**Table 1**. Quantitative results from usability testing[1].

---

**Set of Sonifications and the Rationales**

**Phishing**:

*1. Casting a Fishing Reel.* Users should recognize it as a fishing reel and then connect that with a phishing attack.

*2. Breaking Glass, and 3. Opening a Rust Door.* Phishing attacks often involve attempts to steal information, which is analogous to burglary.

**Malvertising:**

1. *Dropping a Bomb*, *2. Sounding a Siren.* Can wreak havoc on one's computer.

3. *Pouring Water into a Container.* An analogous to the process of downloading a file.

**Form-Filling:**

1. *Typing on a Keyboard.* A component of filling out an online form.

2. *Bubbling Water.* An analogous to an ongoing process such as filling out a form.

3. *Playing a Slot Machine.* The threat during form-filling is exposing sensitive and expensive information.

---

## Case Study

The study involved five participants who were visually impaired, in the age range of 20-49. This sample size is considered sufficient in identifying most system's usability problems [3] including usability studies with users who were visually impaired [2]. Three of the participants were males and four were employed. In terms of education, one had a Master's degree, three had Bachelor's degrees, and one had a high school diploma. They rated their uses of screen readers from Good (2) to Very Good (3). These participants were randomly recruited from a pool of 20 users who are workers and students at a special purpose school for students who are blind or visually impaired and from a state rehabilitation agency for individuals who are blind. All participants reported being blind using screen readers such as: JAWS, Window-Eyes, VoiceOver, NVDA, System Access, and Talkback for Android. See additional demographic information in [5].

Sonifications were created that convey a) the concept or meaning of associated cyber-security threats, and b) the consequences of cyber-threats for users. We focused on threats familiar to most users to ensure the exact intention of usability testing and its effectiveness, namely phishing and malvertising. The former refers to an attacker's attempt to trick users into giving the attacker private information (e.g., users' passwords). The latter, malvertising (malware + advertising) refers to attacks in which an attacker attempts to entice users to download files that contain harmful codes. To include a cyber-security threat that would be familiar to users but operated differently, form-filling of an online form was selected as the potential to expose private information.

*Usability Testing and Participant Tasks*

The usability testing was accomplished in four phases:

**Phase I)** During the first phase of testing, participants listened to the sonifications, which were superimposed on a background of the screen reader JAWS' output. After listening to a given sonification, participants reported which cyber-security threat they thought that the sonification was meant to convey followed by a rationale for their judgment. Participants then rated the sonification's pleasantness, urgency, and conspicuity via 5-point Likert scales (e.g., 1 = extremely unpleasant, 5 = extremely pleasant).

**Phase II)** Similar to Phase I, during the second phase of testing, participants listened to the set of 3 sonifications associated with each cyber-security threat. For each set, participants were asked to choose the sonification that best represented the intended cyber-security threat and to provide a rationale for the choice.

**Phase III)** During the third phase of testing, participants listened to each possible pairing of their choices for the best sonifications of each cyber-security threat. For each pair, participants rated the discriminability of the sonifications on the 5-point Likert scale. During this phase of testing, sonifications were not superimposed onto a JAWS output background. During this phase of testing, participants judged how similar or dissimilar were the sonifications.

---

[1]For sonification and sounds, please visit the following webpage:
http://www.myweb.ttu.edu/asiamina/SonificationSounds.html

**Phase IV**) During the fourth stage of testing, participants again listened to each of the sonifications, which were superimposed on a background of JAWS output. After each sonification, participants were asked whether they remembered what cyber-security threat that sonification was meant to convey. If they answered "yes," participants were asked to report what they thought was the intended cyber-security threat.

*Chosen Earcon-Based Sonification Approach*
The *representational* sonification was chosen mainly for two reasons. First, we wanted to create sounds that could convey their *intended meanings* with *little-to-no* user training. Second, we hypothesized that, to reduce their cognitive loads, individuals with visual impairments heavily utilize natural sounds.

Each project team member independently searched online sound repositories (e.g., www.sounddogs.com) for natural sounds representing phishing, malvertising, and form-filling. Approximately sixty percent more candidate sounds than needed were chosen. Each team member selected and ranked three sounds considered to be the best potential sonification for each cyber-security threat. The highly ranked sonifications for each threat were usability tested in order to investigate whether users would i) hear the indicator when triggered, ii) identify what cyber-security threat a given indicator was meant to convey, and iii) react appropriately to the indicator. Figures 1 – 3 illustrate the sonifications for some of the security threats.

*Results and Discussion*
Table 1 provides a summary of the major quantitative results from the usability testing. Review of the table suggests the following.

- Users correctly identified the cyber-security threat associated with certain sonifications. For example, 80% of users correctly identified that the "casting a fishing reel" sonification warned of a phishing attack. This finding suggests that it is possible to develop sonified cyber-security threat indicators that users intuitively understand.
- Users failed to correctly identify the cyber-security threat associated with certain other sonifications. However, once told the intended sonification-threat mapping, users were able to remember that pairing when tested later. This finding suggests that users' abilities to recognize that a given sonification warns of a given attack could increase as they use such a system or with training.
- There did not seem to be a relation (positive or negative) between pleasantness and the likelihood that 1) the sonification would be correctly identified without training, 2) the sonification's intended meaning would be correctly remembered at the end of the testing session or 3) the sonification would be picked as the best sonification for that cyber-security threat type.

## Conclusions and Next Steps
The results of initial usability testing were promising in that it is possible to develop sonified cyber-security threat indicators that users intuitively understand. Future research should explore ways to optimize various facets of the sonification development process. For example, the process of finding and selecting candidate sonifications was cumbersome; it would be advantageous to develop ways to fully or partially automate that process.

## Acknowledgment
This work is supported by National Science Foundation under award number CNS-1347521. Thanks to Yuanlin Zhang, Fethi Inan, Thomas Hughes, John Rose, Miriam Armstrong, and Tim Salau for the discussion and conducting the usability testing sessions.

## References

1. Meera M. Blattner, Denise A. Sumikawa, Robert M. Greenberg. 1989. Earcons and icons: Their structure and common design principles. *Human–Computer Interaction*, 4(1): 11–44.

2. Elaine Gerber. 2002. Surfing by ear: Usability concerns of computer users who are blind or visually impaired. *Access World*, pages 38–43.

3. Robert A. Virzi. 1992. Refining the test phase of usability evaluation: How many subjects is enough? *Human Factors*, 34:457–468.

4. Thomas Hermann, Andy Hunt, John G. Neuhoff. 2011. *The Sonification Handbook*. Logos Verlag.

5. Fethi Inan, Akbar Siami Namin, Keith S. Jones, Rona Pogrund, 2016, Perception of Cybersecurity Risks for Internet Users Who Are Visually Impaired, *Journal of Educational Technology and Society*. (ISSN 1436-4522).

6. Petrie, Fisher, ONeill, Fisher, Di Segni Y. 2001. Deliverable 2.1: Report on user requirements of mainstream readers and print disabled readers.

7. Romisa R. Ghahari, Mexhid Ferati, Tao Yang, Davide Bolchini. 2012. Back navigation shortcuts for screen reader users. In Proceedings of the 14th international ACM SIGACCESS conference on Computers and accessibility, pages 1–8.

8. Jonathan Lazar, Allen Aaron, Jason Kleinman, Chris Malarkey. 2007. What frustrates screen reader users on the web: A study of 100 blind users. International Journal of human-computer interaction, 22(3): 247–269.

9. Internet Security Threat Report (ISTR 20). 2015 Volume 20, Symantec Annual Report.

10. A. Siami Namin, R. Hewett, K.S. Jones, and R. Pogrund. 2016. Sonifying Internet Security Threats, ACM International Conference on Computer-Human Interactions, Late-Breaking Work (ACM CHI'16), San Jose, California, USA.